\documentclass[runningheads]{llncs}
\usepackage[T1]{fontenc}
\usepackage{graphicx}
\usepackage{xcolor}
\usepackage{todonotes}
\usepackage{listings}

\definecolor{codegreen}{rgb}{0,0.6,0}
\definecolor{codegray}{rgb}{0.5,0.5,0.5}
\definecolor{codepurple}{rgb}{0.58,0,0.82}
\definecolor{backcolour}{rgb}{0.95,0.95,0.92}

\lstdefinestyle{mystyle}{
    backgroundcolor=\color{backcolour},   
    commentstyle=\color{codegreen},
    keywordstyle=\color{magenta},
    numberstyle=\tiny\color{codegray},
    stringstyle=\color{codepurple},
    basicstyle=\ttfamily\footnotesize\tiny,
    breakatwhitespace=false,         
    breaklines=true,                 
    captionpos=b,                    
    keepspaces=true,                 
    numbers=left,                    
    numbersep=5pt,                  
    showspaces=false,                
    showstringspaces=false,
    showtabs=false,                  
    tabsize=2
}

\usepackage{hyperref}
\hypersetup{
    colorlinks=true,
    linkcolor=blue,
    filecolor=magenta,      
    urlcolor=teal,
    }

\usepackage{tikz}
\usetikzlibrary{shapes, arrows, positioning}

\begin{document}
%
\title{Federated SPARQL querying for genomic variant functional annotation}

%
%
\author{\small Alexandrina Bodrug-Schepers\inst{1}\tiny\orcidID{0009-0006-3873-102X}\and\small Romain Bourcier\inst{1}\tiny\orcidID{0000-0002-6506-4019} 
\and\small Richard Redon\inst{1}\tiny\orcidID{0000-0001-7751-2280} \and\small Alban Gaignard\inst{1,2}\tiny\orcidID{0000-0002-3597-8557}}
\authorrunning{A. Bodrug-Schepers et al.}
%
\institute{Nantes Université, CHU Nantes, CNRS, INSERM, l’institut du thorax, F-44000 Nantes, France \and
IFB-core, Institut Français de Bioinformatique (IFB), CNRS, INSERM, INRAE, CEA, 91057 Evry, France}
\maketitle              
\begin{abstract}
Sensitive health data should preferentially be analysed on site. In typical bioinformatics workflows, public databases are duplicated and used by specialised tools to enrich the local datasets. In the case of genomic variation data, this process is called variant annotation. In this session we demonstrate variant annotation using federated SPARQL queries. We first overview how clinico-genomic data can be modelled as a knowledge graph (KG), leveraging state-of-the-art biomedical ontologies. We then perform variant annotation by querying UniprotKB, a massive curated KG for gene and proteins. Our approach avoids public data duplication while maintaining genomic data on site and aligning it with FAIR principles. Our use-case is based on the ICAN project, a research program aimed at studying the physiopathology of cerebral berry aneurysms.

\end{abstract}
%



\section{Introduction} \label{intro}

Clinico-genomic data are common in biomedical research investigating genetic factors. The rise of bioinformatics has improved the ability to produce and analyse these massive datasets~\cite{gnomad}. In practice, locally generated data are enriched through duplication of public biological databases. This approach is suitable for small, well-defined datasets and is preferred when data is sensitive and should remain on-site. 
Massive genomics knowledge bases contain key information such as pathogenicity scores or general population allele frequencies to distinguish between rare and common variants~\cite{gnomad}. 
Their duplication is costly in terms of storage and computational resources. Moreover, the variant annotation pipelines must be regularly re-run to remain aligned with state-of-the-art knowledge.

In this work we demonstrate genomic variant annotation using federated SPARQL querying.
First, we briefly explain our method for the transformation of typical tabulated formats, \textit{i.e.} clinical relational databases and Variant Calling Format (VCF) files, into a KG leveraging biomedical ontologies. We then discuss the benefits: (1) alignment with FAIR principles, (2) heterogeneous data integration and (3) interoperability with biomedical KGs. We finally showcase this third aspect with live federated SPARQL queries. Our approach is novel as it allows semantic  interoperability of private genomic data with publicly available knowledge graphs while avoiding two common challenges in more conventional genomic analyses workflows \textit{i.e.} public databases duplication and private genomic data sharing.

The demo scenarios annotate genomic variants, hereafter called just variants, using Wikidata, a generalistic trusted KG and UniprotKB~\cite{uniprot}, a biomedical KG with a focus on proteins. Our use-case is based on the ICAN project~\cite{ican2017_shift}, a research program aimed at studying the physiopathology of cerebral berry aneurysms (CBA). CBAs are abnormal bulges in the blood vessels of the brain that pose significant health risks. 

\begin{figure}
\begin{center}
\includegraphics[width=1\textwidth]{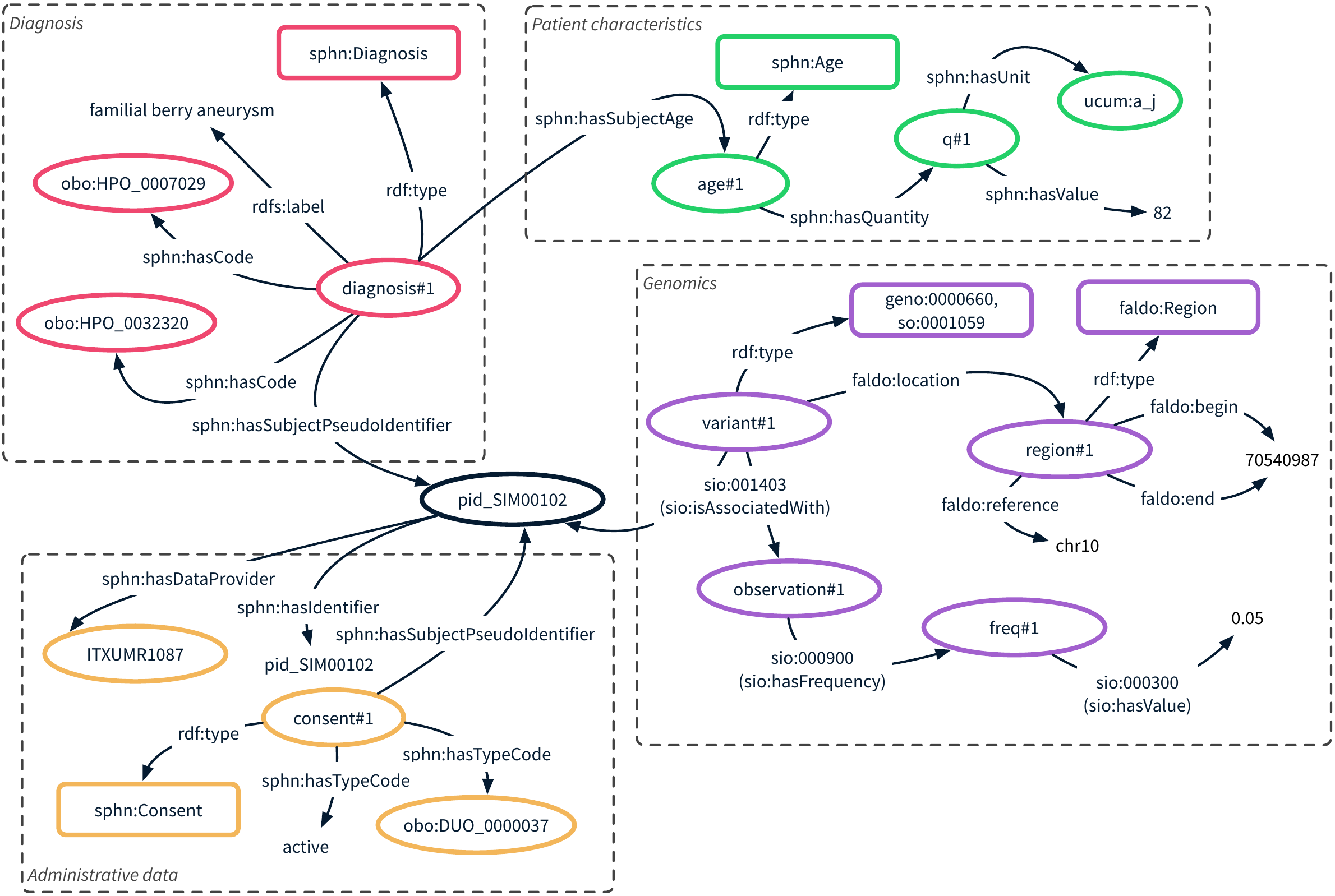}    
\caption{Example of a synthetic individual modelled in CGen-KG}
\end{center}
\label{fig_model}
\end{figure}

\vspace{-1cm}
\section{From biomedical data to biomedical KG} \label{icandata2kg}

\subsubsection{Synthetic patients.}
We simulated 3,000 individuals with 45 features as in~\cite{parcours} and >70,000 variants each, located in genomic regions coding for proteins (genes). To make our dataset as realistic as possible, synthetic data were generated according to observations made on 800 individuals from the ICAN projet. All individuals were diagnosed with at least one CBA. The simulated clinical data is similar in content type, structure, and probability distributions according to phenotypic criteria, without considering feature dependencies to limit re-identification risks. We used Mozza\textsuperscript{1}\footnotetext[1]{https://github.com/genostats/Mozza} to simulate synthetic haplotypes from variants of individuals with CBA. As a precaution, we excluded potentially identifiable rare variants (allele frequencies below 5\%).
\vspace*{-\baselineskip}
\subsubsection{Implementation.} 
The demo system was deployed on an 8 vCPU, 8 GB RAM, 20 GB storage virtual machine and includes: (i) an Apache Jena Fuseki triple store containing the RDF data; and (ii) a web interface built with Streamlit. The resulting publicly accessible web app, called CGen-KG\textsuperscript{2}\footnotetext[2]{https://cgen-kg-ica.bird.glicid.fr/}, contains 43,063,591 triples. he demo code is publicly available\textsuperscript{3}\footnotetext[3]{https://gitlab.univ-nantes.fr/bodrug-a/demo-aggrvarkg}.
\vspace*{-\baselineskip}
\subsubsection{Data modeling.} We used state-of-the-art biomedical ontologies to model our synthetic data (Figure\ref{fig_model}). The modelling is a continuation of our previous works~\cite{sembeacons}. We used the \href{https://www.biomedit.ch/rdf/sphn-schema/sphn/2025/2#}{SPHN} ontology, \href{https://ontobee.org/ontology/HP}{HPO} and \href{https://ontobee.org/ontology/DUO}{DUO} to model individuals' clinical features. We modelled individuals with CBA familial forms; their age at diagnosis, as well as their consent. We used \href{https://ontobee.org/ontology/GENO}{GENO}, \href{https://ontobee.org/ontology/SO}{SO} and \href{http://biohackathon.org/resource/faldo}{FALDO} to represent variants, their alleles and locations. We used \href{https://ontobee.org/ontology/SIO}{SIO} as a semantic connector to link variants to individuals and to their allele frequencies, a key aspect in genomics analysis. In summary, a variant was linked to three main larger concepts 1) its genomic location, 2) individuals carrying it and 3) phenotype related allele frequencies (see relationships of 'variant\#1' in Figure \ref{fig_model}).

\section{Live demo} \label{fedquery}
During this session we annotate variants found in GGen-KG with Gene Ontology (GO) terms. GO terms are used to describe molecular function, cellular localisation and biological processes of gene products \textit{e.g.} proteins. 

{\bf \em 1. KG statistics.} The user selects the "Statistics" tab to view the SPARQL queries and the dynamically generated plots. CGen-KG content and parts of UniprotKB and Wikidata used for variant annotation in the next step are viewed. 

{\bf \em 2. Variant annotation.} The user selects the "Annotation" tab where we showcase federated SPARQL queries for realistic bioinformatics scenarios. 

In \textit{Scenario 1.}, the user is investigating a variant. They wish to know the following information: (i) in which gene is the variant located, (ii) which protein does the variant code for, (iii) which GO terms can we associate to the variant, (iv) which diagnoses are associated with the variant. 

\begin{figure}
\begin{center}
\includegraphics[width=1\textwidth]{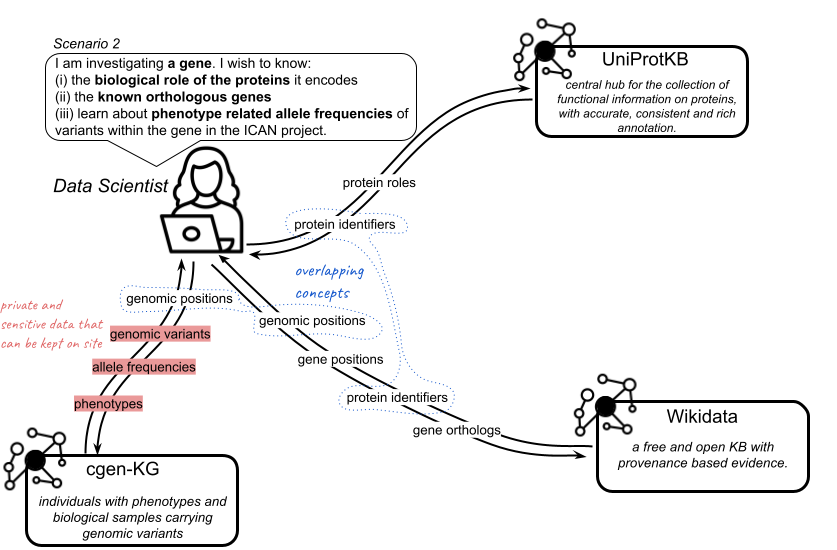}    
\caption{End-to-end motivational question}
\end{center}
\label{fig_motivationalQ}
\end{figure}

In \textit{Scenario 2.}, the user is investigating a gene. They wish to know (i) the biological role of the proteins it encodes, (ii) retrieve a list of known orthologous genes and (iii) learn about allele frequencies of variants in this gene. The knowledge sources to answer these questions is detailed in Figure \ref{fig_motivationalQ}.

Finally, in \textit{Scenario 3.}, the user wants to create a subset of variants present in the ICAN population that affect proteins involved in blood vessel formation. They use the \href{https://www.ebi.ac.uk/QuickGO/term/GO:0001525}{GO:0001525} term. The query is detailed in Figure \ref{fig_query}. 




\begin{figure}
\begin{lstlisting}[language=SPARQL]
SELECT ?variantId ?variantStart ?variantEnd ?variantChromosome 
?geneStart ?geneEnd ?geneChromosome ?geneAssembly 
?proteinName ?proteinID2 ?goTerm 

WHERE {
  SERVICE <https://sparql.uniprot.org/sparql> { # UniprotKB
    ?protein a up:Protein ;
    up:organism taxon:9606 ;  # taxon term for Homo sapiens   
    up:classifiedWith GO:0001525 . # GO term for blood vessel formation
    ?protein up:mnemonic ?proteinName .
  }
  BIND(SUBSTR(STR(?protein), STRLEN(STR(up:)) + 4) AS ?proteinID2) . 
  SERVICE <https://query.wikidata.org/sparql> { # Wikidata
    ?wp wdt:P352 ?proteinID2 ; wdt:P702 ?wg . # gene
    ?wg wdp:P644 ?wgss ; wdp:P645 ?wgse . # gene location
    ?wgss wdps:P644 ?geneStart ; wdpq:P1057/wdt:P1813 ?geneChromosome ; wdpq:P659/wdt:P2576 ?geneAssembly .
   	?wgse wdps:P645 ?geneEnd ; wdpq:P1057/wdt:P1813 ?geneChromosome ; wdpq:P659/wdt:P2576 ?geneAssembly . 
    FILTER(STR(?geneAssembly) = 'hg38') # gene assembly
    FILTER(STR(?geneChromosome) = "19") # gene chromosome
  }
  ?variant a so:0001059 . # CGen-KG variant
  ?variant sio:000671/sio:000300 ?variantId .
  ?variant faldo:location/faldo:begin/faldo:position ?variantStart .
  ?variant faldo:location/faldo:end/faldo:position ?variantEnd .
  ?variant faldo:location/faldo:reference/rdf:label ?variantChromosome .

  FILTER(STR(?geneChromosome) = STR(?variantChromosome))   # CGen-KG variant location same as and Wikidata gene location
  FILTER((xsd:integer(?variantStart) >= xsd:integer(?geneStart)
          && xsd:integer(?variantStart) <= xsd:integer(?geneEnd)) 
     || (xsd:integer(?variantEnd) >= xsd:integer(?geneStart) 
         && xsd:integer(?variantEnd) <= xsd:integer(?geneEnd))) # variant within the gene
}
\end{lstlisting}
\caption{Federated SPARQL query linking public protein knowledge and local variants} \label{fig_query}
\end{figure}

\section{Conclusion} \label{conclusion}
Existing biomedical data integrated as KGs had goals different to ours and consequently mixed private and public datasets, duplicated databases, shared patient data and built in-house ontologies~\cite{prasanna2025}. In contrast, our approach ensures all the following aspects: 1) mapping of a biomedical dataset to state-of-the-art ontologies, 2) avoiding database duplication by allowing federated querying, and 3) keeping private and sensitive data on-site. This organisation enables a data discovery step upstream of more conventional bioinformatics pipelines \textit{i.e.} one can explore existing annotations before committing to full scale analyses. As future works, we aim at better preserving security concerns in the context of biomedical SPARQL endpoint federations (ANR-25-CE23-7852 Safe-KG grant). We consider the main limitation of variant annotation with federated SPARQL querying to be the lack of reproducibility as the hosted RDF dataset may evolve during time. However, solutions have already been described~\cite{glenda}. To conclude, with this demo we wish to send the following message: using semantic web technologies to align human genomic data to FAIR principles promotes discoverability, creates new opportunities and does not require sensitive data sharing.




\begin{credits}
\subsubsection{\ackname} \tiny This work was funded by the French government, through the National Research Agency (ANR), under the “France 2030” program with reference ANR-22-PESN-0008. We are most grateful to the Genomics Core Facility GenoA, member of Biogenouest and France Genomique and to the Bioinformatics Core Facility BiRD, member of Biogenouest and Institut Français de Bioinformatique (IFB) (ANR-11-INBS-0013) for the use of their resources and their technical support.

\end{credits}
%
%
%
\bibliographystyle{splncs04_lite}
\bibliography{papers.bib}

\end{document}